\documentstyle[12pt,epsf]{article}

\newcommand{\be}{\begin{equation}}
\newcommand{\ee}{\end{equation}}
\newcommand{\bea}{\begin{eqnarray}}
\newcommand{\eea}{\end{eqnarray}}
\newcommand{\ba}{\begin{array}}
\newcommand{\ea}{\end{array}}

\newcommand{\rk}{{\rm rank}\,}
\newcommand{\diag}{{\rm diag}}
\newcommand{\const}{{\rm const.}}
\newcommand{\rv}{\vec{r}}
\newcommand{\w}{\vec{\omega}}

\newcommand{\journal}[4]{{\rm #1} {\bf #2} (19#3) #4}
\newcommand{\NP}{\journal{Nucl. Phys.}}
\newcommand{\PL}{\journal{Phys. Lett.}}
\newcommand{\PRL}{\journal{Phys. Rev. Lett.}}
\newcommand{\NPPS}{\journal{Nucl. Phys. Proc. Suppl.}}
\newcommand{\MPCPS}{\journal{Math. Proc. Camb. Phil. Soc.}}

\makeatletter
  
  \@addtoreset{equation}{section}
\makeatother

\begin{document}

\baselineskip=18pt

%
%
\begin{titlepage}

\begin{flushright}
OU-HET 275 \\
hep-th/9707258 \\
July 1997
\end{flushright}
\bigskip
\bigskip

\begin{center}
{\Large \bf
Higgs Branch of $N=2$ SQCD\\
and\\
$M$ theory Branes\\
}
\bigskip
Toshio Nakatsu, Kazutoshi Ohta, Takashi Yokono and Yuhsuke Yoshida\\
\bigskip
{\small \it
Department of Physics,\\
Graduate School of Science, Osaka University,\\
Toyonaka, Osaka 560, JAPAN
}
\end{center}
\bigskip
\bigskip
\begin{abstract}
Higgs branch of $N=2$ SQCD is studied from the $M$ theory viewpoint. 
With a differential geometrical proof of the $s$-rule besides 
an investigation on the global symmetry of $M$ theory brane 
configurations, an exact description of the baryonic and 
non-baryonic branches in terms of $M$ theory is presented. 
The baryonic branch root is also studied. 
The ``electric'' and ``magnetic'' descriptions of the root 
are shown to be related with each other by the brane exchange 
in $M$ theory.

\end{abstract}

\end{titlepage}

%
%

\section{Introduction}

                    It has brought about many novel results 
to analyze supersymmetric field theory
as the effective world volume theory of branes 
in superstring theory. 
Simultaneously, the world volume theory of branes 
provides a very useful tool for our understanding of  
the non-perturbative dynamics of superstring. 
Recently  $M$ theoretical description 
of four-dimensional $N=2$ supersymmetric gauge theory is 
proposed by Witten \cite{W1}. 
The mysterious hyper-elliptic curve, 
used for the exact solution of the Coulomb branch 
of $N=2$ supersymmetric QCD (SQCD), 
now becomes a part of a $M$ theory fivebrane.

                      On the other hand, 
the moduli space of vacua of $N=2$ SQCD is investigated 
\cite{APS} in detail including various Higgs branches 
from a field theoretical point of view. 
It was argued that the root of baryonic branch,  
where the Higgs branch and the Coulomb branch 
come in contact with each other, 
will play a crucial role, 
after breaking $N=2$ supersymmetry to $N=1$, 
in understanding the so-called $N=1$ non-Abelian duality 
\cite{Seiberg,IS}. 
However, $M$ theoretical interpretation of the Higgs branch 
and a further understanding of the duality in this direction 
are still not clear.

                     In this article 
we study the Higgs branch of $N=2$ SQCD from the view of 
$M$ theory. An exact description of the baryonic 
and non-baryonic branches in terms of $M$ theory 
will be presented. We also study the baryonic branch root. 
Its two different descriptions \cite{APS} in field theory 
will be shown to be related with each other in $M$ theory 
by exchanging parts of branes. 
We give a brief review 
on the moduli space of the Higgs branch of $N=2$ SQCD 
in Section 2.  
It is introduced as a hyper-K\"ahler quotient space 
and classified into two parts according to the colour 
symmetry breaking patterns;  
baryonic branch and non-baryonic branch. 
The flavor symmetry group can act on the moduli space. 
The residual global symmetry of each class is described 
as its stabilizer. 
The duality between the baryonic branches 
of $SU(N_c)$ and $SU(N_f-N_c)$ gauge theories is shown 
by using an algebraic geometrical description of these 
moduli spaces.

             In Section 3 we consider the $M$ theory description 
of $N=2$ SQCD systems.  The Seiberg-Witten hyper-elliptic curve 
becomes a part of a fivebrane and is embedded into the multi 
Taub-NUT space. This embedding of the curve is studied in detail 
from the differential geometrical viewpoint. 
The formulae obtained there are used for a proof of the $s$-rule  
\cite{HW}. An exact description of the baryonic and non-baryonic 
branches is given in terms of the brane configuration.

              In the last section we consider a family of 
finite (scale-invariant) theory which is reducible, 
by taking a double scaling at the weak coupling limit, 
to the baryonic branch root of asymptotically free (AF) theory 
and provides the ``electric'' description of the baryonic branch 
root. By changing the value of the bare coupling constant 
$\tau$ we find that the brane exchange occurs possibly 
on a semi-circle with radius $1$ in the upper half $\tau$-plane. 
On this semi-circle, due to the brane exchange, 
the original brane configuration becomes a dual one. 
This dual configuration provides another description of the 
baryonic branch root. 
One can expect that these two configurations give 
the $N=1$ non-Abelian dual brane configurations \cite{EGK} 
after rotating a part of brane \cite{Barbon,HOO,W2}.

%
%

\section{Review of Higgs Branch}

\subsection{Baryonic Branch and Non-Baryonic Branch}

         In this section we briefly review the Higgs branch
of $N=2$ supersymmetric QCD to study a connection
with brane configurations.

      Consider $N=2$ $SU(N_c)$ supersymmetric QCD
with $N_f$ flavors.
We will treat mainly the case of
$N_c \leq N_f < 2N_c$ in this article.
The generalization to other cases is straightforward.
In this SUSY theory
there exist a $N=1$ chiral multiplet $\Phi$
in the adjoint representation of the gauge group
$SU(N_c)$,
and $N=1$ chiral multiplets $Q$ and $\tilde{Q}$
which respectively belong to
the fundamental and anti-fundamental
representations of the gauge group besides
the flavor group $U(N_f)$.
These matter chiral multiplets,
regarding them as $SU(2)_R$ doublets,
give $N=2$ hypermultiplets.
It becomes convenient to write 
$Q$ as a complex $N_c\times N_f$ matrix and
$\tilde{Q}$ as a complex $N_f\times N_c$ matrix.

                   The classical vacua of the theory
are described by the F- and D-flat conditions.
Among these vacuum configurations
we shall consider the pure Higgs branch, that is,
the branch characterized by $\Phi=0$.
In this branch the F- and D-flat conditions are
read as
\bea
Q\tilde{Q} &=& \rho{\bf 1}_{N_c}, \nonumber \\
QQ^{\dagger}-\tilde{Q}^{\dagger}\tilde{Q} &=&  \nu{\bf 1}_{N_c},
\label{flat eqs}
\eea
where ${\bf 1}_{N_c}$ is the  $N_c\times N_c$ unit matrix.
$\rho$ and $\nu$ are respectively  complex and real numbers
which we call the Fayet-Iliopoulos (FI) parameters.
The consideration on the Higgs branch is separated
into two cases whether the FI terms vanish or not.
The former case is non-baryonic and the latter is baryonic.

                 Let us first examine the baryonic branch.
Regarding $Q,\tilde{Q}$ as $N=2$ hypermultiplets $Q_{\dot{A}}$
($\dot{A}=\dot{1},\dot{2}$) by the identification
\be
Q_{\dot{1}}\equiv Q,~~~Q_{\dot{2}}\equiv \tilde{Q}^{\dagger},
\ee
it is possible to rewrite eqs. (\ref{flat eqs}) in
the $SU(2)_R$-covariant fashion
\be
\mu_{\dot{A}\dot{B}}\equiv
Q_{(\dot{A}}\bar{Q}_{\dot{B})}=\zeta_{\dot{A}\dot{B}}{\bf 1}_{N_c},
\label{eq of vacua}
\ee
where
$\bar{Q}_{\dot{A}}\equiv\epsilon_{\dot{A}\dot{B}}Q_{\dot{B}}^{\dag}$.
$\zeta_{\dot{A}\dot{B}}$ is the $SU(2)_R$-covariant form of
the FI parameters
\bea
{\zeta^{\dot{A}}}_{\dot{B}}
 &=& \epsilon^{\dot{A}\dot{C}}\zeta_{\dot{C}\dot{B}} \nonumber \\
 &=& \left(
\ba{cc}
\nu/2 & \rho^* \\
\rho  & -\nu/2
\ea
\right).
\eea
Notice that
$\mu_{\dot{A}\dot{B}}=Q_{(\dot{A}}\bar{Q}_{\dot{B})}$
is the hyper-K\"ahler momentum map of
the $U(N_c)$ action,
\bea
Q_{\dot{A}}\mapsto Q_{\dot{A}}'=gQ_{\dot{A}}~~~~~~
g \in U(N_c).
\label{U(Nc)}
\eea

          Now the moduli space of the baryonic branch is
given by the space of the gauge equivalent classes of the
solutions of vacuum equation (\ref{eq of vacua}).
It has the form
\be
{\cal M}_{\rm b} =
 ({\bf C}\times{\bf R}\backslash (0,0))\times
\frac{\mu_{\dot{A}\dot{B}}^{-1}(\zeta_{\dot{A}\dot{B}})}{SU(N_c)}
\label{M1}
\ee
Notice that
$\det{\zeta^{\dot{A}}}_{\dot{B}}\neq 0$ holds
in the baryonic branch.
By the isomorphism  $SU(N_c) \simeq U(N_c)/U(1)$ we may modify
the quotient in (\ref{M1}) as follows:
\be
{\cal M}_{\rm b} = P
\times\widehat{\cal M}_{\rm b}(\zeta_{\dot{A}\dot{B}}),
\label{M2}
\ee
where
\be
P = ({\bf C}\times{\bf R}\backslash (0,0))\times U(1),
\label{P}
\ee
\be
\widehat{\cal M}_{\rm b}(\zeta_{\dot{A}\dot{B}})
=\frac{ \mu_{\dot{A}\dot{B}}^{-1}(\zeta_{\dot{A}\dot{B}})}{U(N_c)}.
\label{submoduli}
\ee
$\widehat{\cal M}_{\rm b}(\zeta_{\dot{A}\dot{B}})$ is
the hyper-K\"ahler quotient of the $U(N_c)$-action.
One can also regard
any point in $P$, $(\rho,\nu,e^{i\phi})$,
as a vev of one hypermultiplet.
As for the dimensionality of the moduli space
it turns out, simply counting up the degrees of freedom,
to be
\be
\dim_{\bf R}{\cal M}_{\rm b}=4(N_c(N_f-N_c)+1).
\ee

       To study the moduli space it is useful to
introduce a complex structure of
$\widehat{\cal M}_{\rm b}(\zeta_{\dot{A}\dot{B}})$.
With a given complex structure one can split the hyper-K\"ahler
momentum map $\mu_{\dot{A}\dot{B}}$ into
a complex part $\mu_{\bf C}$ and a real part $\mu_{\bf R}$.
Actually we will take the complex structure such that
eq. (\ref{eq of vacua}) acquires the form
\bea
\mu_{\bf C} &\equiv&
H_{\dot{1}}H_{\dot{2}}^{\dagger} = 0
\label{muC} \\
\mu_{\bf R} &\equiv&
H_{\dot{1}}H_{\dot{1}}^{\dagger}-H_{\dot{2}}H_{\dot{2}}^{\dagger}
 = \eta{\bf 1}_{N_c}.
\label{muR}
\eea
The FI parameters $\zeta_{\dot{A}\dot{B}}$
are rotated by $SU(2)_R$ equal to zero
except the only one component,
which is denoted by $\eta$.
Here $\eta \equiv\sqrt{|2\rho|^2+\nu^2}$ is  positive
for the baryonic branch. And $H_{\dot{A}}$
is the corresponding $SU(2)_{R}$-rotated form of $Q_{\dot{A}}$.
The hyper-K\"ahler quotient is regarded as
\be
\widehat{\cal M}_{\rm b}(\zeta_{\dot{A}\dot{B}})
\simeq
\frac{\mu_{\bf C}^{-1}(0)\cap\mu_{\bf R}^{-1}(\eta)}{U(N_c)}. 
\label{Mhat}
\ee

                 The positivity of $\eta$ ensures the following
stability condition for any solution of eqs.(\ref{muC}) and
(\ref{muR})
\be
\rk H_{\dot{1}}=N_c,
\label{stability}
\ee
from which we can conclude that
$\widehat{\cal M}_{\rm b}(\zeta_{\dot{A}\dot{B}})$ is
a {\it smooth} hyper-K\"ahler manifold.
This is because, though any possible singularity
in $\widehat{\cal M}_{\rm b}(\zeta_{\dot{A}\dot{B}})$,
if it exists, originates in a solution of
$\mu_{\dot{A}\dot{B}}=\zeta_{\dot{A}\dot{B}}$
on which $U(N_c)$ does not act transitively,
the stability condition (\ref{stability}) can be shown to
prohibit such a fixed point.

                     The flavor group
$U(N_f)$ acts on the hyper-K\"ahler quotient
$\widehat{\cal M}_{\rm b}(\zeta_{\dot{A}\dot{B}})$.
Its action is given by
\be
H_{\dot{A}} \mapsto H_{\dot{A}}'=H_{\dot{A}}f ~~~~~~f \in U(N_f).
\label{U(Nf)}
\ee
Let us consider the stabilizer of this $U(N_f)$-action
at a generic point.
We first remark that,
by taking an appropriate element of
$U(N_f)$,
{\it any} point of $\widehat{\cal M}_{\rm b}(\zeta_{\dot{A}\dot{B}})$
is transformed into the following normal form
\bea
H_{\dot{1}}' &=& \left(
\ba{ccc}
{\bf D}+i\sqrt{\eta}{\bf 1}_{N_f-N_c} & & \phantom{{\bf D}}\\
 & i\sqrt{\eta}{\bf 1}_{2N_c-N_f} &
\ea
\right),
\nonumber  \\
H_{\dot{2}}' &=& \left(
\ba{ccc}
\phantom{{\bf D}+i\sqrt{\eta}{\bf 1}_{N_f-N_c}} & & {\bf D} \\
 & \phantom{i\sqrt{\eta}{\bf 1}_{2N_c-N_f}} &
\ea
\right),
\label{nfb}
\eea
where ${\bf D}=\diag(d_{1},\ldots,d_{N_f-N_c})$ with
any $d_j$ being  a non-negative real number.
The equalities in eqs.(\ref{nfb})
are understood as the equalities mod the $U(N_c)$-action.
{}From this normal form we can see that
the stabilizer $\hat{H}$ is generically isomorphic to
$U(2N_c-N_f) \times U(1)^{N_f-N_c}$ $(\subset U(N_f))$.
However, this $\hat{H}$ is  not
equal to the stabilizer of the $U(N_f)$-action
on the whole moduli space ${\cal M}_{\rm b}$.
That is, 
one of the U(1) factors in $\hat{H}$ acts also on
the additional hypermultiplet $P$ and moves its vev.
This is simply because the whole moduli space
${\cal M}_{\rm b}$ is originally introduced as the direct
product of the space of the FI parameters and the
$SU(N_c)$-quotient while we are studying
it by emphasizing its hyper-K\"ahler structure.
Therefore the net stabilizer $H$ of
the $U(N_f)$-action on the whole moduli space
${\cal M}_{\rm b}$ is $SU(2N_c-N_f) \times U(1)^{N_f-N_c}$.

               The space of vevs $d_j$, which we denote
by $X_{\rm b}(\zeta_{\dot{A}\dot{B}})$,
is a real manifold admitting singularities due to the enhancement
of the stabilizer.
This symmetry enhancement occurs when some of $d_j$ equal to zero
or coincide with one another.
As for the dimensionality it satisfies the following relation
\be
\dim U(N_f)/\hat{H} + \dim_{\bf R}X_{\rm b}(\eta)
 = \dim_{\bf R}\widehat{\cal M}_{\rm b}(\eta).
\ee

             Next, let us consider
the case that the FI parameters vanish, that is,
$\eta=0$. This branch is called the non-baryonic branch.
Consider the hyper-K\"ahler quotient
\be
{\cal M}_{\rm nb}=
\frac{\mu_{\dot{A}\dot{B}}^{-1}(0)}{U(N_c)}.
\ee
This is a singular manifold but possible to give
a stratification by the colour symmetry breaking patterns.
For the breaking pattern,
$SU(N_c) \rightarrow  SU(N_c-r)$
$(0 \leq r \leq [N_f/2])$,
we can introduce a stratum
${\cal M}^{(r)}_{\rm nb}$ as the space of
the gauge equivalent classes of the solutions
of $\mu_{\dot{A}\dot{B}}=0$ satisfying
the condition, $\rk Q_{\dot{A}}=r$ :
\be
{\cal M}_{\rm nb}=
\cup_{r=0}^{[N_f/2]}{\cal M}^{(r)}_{\rm nb}.
\label{strata of nb}
\ee
Each stratum ${\cal M}^{(r)}_{\rm nb}$
can be considered as a hyper-K\"ahler manifold
of dimensions $4r(N_f-r)$ and called the
$r$-th non-baryonic branch.
By the $U(N_f)$-action
any point of the $r$-th non-baryonic branch
can be transformed into the following normal
form :
\bea
Q_{\dot{1}}' &=& \left(
\ba{ccc}
{\bf D}^{(r)} &\phantom{{\bf 0}}   &  \phantom{{\bf D}^{(r)}}  \\
              &                     &
\ea
\right),\nonumber \\
Q_{\dot{2}}' &=& \left(
\ba{ccc}
\phantom{{\bf D}^{(r)}} &  \phantom{{\bf 0}} & {\bf D}^{(r)}  \\
                        &                   &
\ea
\right),
\label{nfnb}
\eea
where
${\bf D}^{(r)}=\diag(d_{1},\ldots,d_{r})$ with any
$d_j$ being a positive real number.
(The above equalities are also understood mod the
$U(N_c)$-action.)
{}From this normal form we can read that
the stabilizer $H$ of the $U(N_f)$-action
is generically isomorphic to
$U(N_f-2r)\times U(1)^r$.
Notice that $U(1)^r$ moves $Q_{\dot{A}}'$
but this transform can be identified with the original one in
${\cal M}^{(r)}_{\rm nb}$ by the $U(N_c)$-action
while $U(N_f-2r)$ does not move $Q_{\dot{A}}'$ at all.
The space of vevs $d_j$, which we denote
by $X_{\rm nb}^{(r)}$,
is a real manifold admitting singularities due to the enhancement
of the stabilizer.
This symmetry enhancement occurs when some of $d_j$
coincide with one another.
As for the dimensionality it satisfies the following relation
\be
\dim U(N_f)/H + \dim_{\bf R}X_{\rm nb}^{(r)}
 = \dim_{\bf R}{\cal M}_{\rm nb}^{(r)}.
\ee

                 At this stage
it might be also convenient to remark
on the relation between the baryonic and non-baryonic branches.
Notice that we can define an inclusion of
$\widehat{{\cal M}}_{\rm b}(\zeta_{\dot{A}\dot{B}})$ into
${\cal M}_{\rm nb}$ by simply taking the limit of
$\zeta_{\dot{A}\dot{B}} \rightarrow 0$.
By comparing normal forms (\ref{nfb}) and (\ref{nfnb})
its image will be
$\cup_{r=0}^{N_f-N_c}{\cal M}^{(r)}_{\rm nb}$.
Since
$\widehat{{\cal M}}_{\rm b}(\zeta_{\dot{A}\dot{B}})$
is a smooth hyper-K\"ahler manifold ($\zeta_{\dot{A}\dot{B}} \neq 0$)
we can say that it is giving a resolution of
$\cup_{r=0}^{N_f-N_c}{\cal M}^{(r)}_{\rm nb}$.

\subsection{Duality of Higgs Branches}

                     Looking at the normal form (\ref{nfb})
one may find that the upper half blocks of these matrices is
identified with the normal form in the moduli space of
$SU(N_f-N_c)$ gauge theory with $N_f$ flavors.
This indicates that there is an
isomorphism among the baryonic branches of $SU(N_c)$
and $SU(N_f-N_c)$ gauge theory with the same $N_f$ flavors.
To confirm this observation,
let us consider the baryonic branch
from an algebraic geometrical viewpoint.

                 Since the positivity of $\eta$
makes the moduli space
$\widehat{\cal M}_{\bf b}(\zeta_{\dot{A}\dot{B}})$
(\ref{submoduli}) smooth,
we must introduce its algebraic counterpart
as a ``regularized'' complex symplectic quotient.
It will be given by the following complex symplectic
quotient regularized by stability condition (\ref{stability}):
\be
{\cal M}^{alg}_{\rm b}(N_c,N_f)\equiv
\frac{\mu_{\bf C}^{-1}(0)^s}{GL(N_c,{\bf C})},
\label{algebraic M}
\ee
where
\be
\mu_{\bf C}^{-1}(0)^s=\left\{H_{\dot{A}}\left|
\ba{l}
\cdot ~  \mu_{\bf C}=0  \\
\cdot ~  \rk H_{\dot{1}}=N_c
\ea
\right.\right\}.
\ee
Notice that $GL(N_c,{\bf C})$ is a complexification of $U(N_c)$
and acts on $H_{\dot{A}}$ as
\be
(H_{\dot{1}},H_{\dot{2}}^{\dagger}) \mapsto
(gH_{\dot{1}},H_{\dot{2}}^{\dagger}g^{-1})
{}~~~~~~ g \in GL(N_c,{\bf C}).
\label{GL(Nc)}
\ee
This $GL(N_c,{\bf C})$-quotient
turns out to define a {\it smooth} algebraic variety.
(Without the stability condition
the complex symplectic quotient
$\mu_{\bf C}^{-1}(0)/GL(N_c,{\bf C})$
have several singularities.)
This  introduction of
the algebraic counterpart of the moduli space
actually makes
the parametrization of
${\cal M}^{alg}_{\rm b}(N_c,N_f)$
very simple. It is described by
\bea
H_{\dot{1}} &=& \left(
\ba{cc}
\phantom{-h_{\dot{2}}h_{\dot{1}}^{\dagger}} & \\
{\bf 1}_{N_c}  & h_{\dot{1}} \\
               &
\ea
\right),
\label{h1}
\\
H_{\dot{2}} &=& \left(
\ba{cc}
                                  & \\
-h_{\dot{2}}h_{\dot{1}}^{\dagger} & h_{\dot{2}} \\
                                  &
\ea
\right),
\label{h2}
\eea
where $h_{\dot{A}}$ are arbitrary
$N_c \times (N_f-N_c)$ complex matrices and give
the holomorphic coordinates of
${\cal M}^{alg}_{\rm b}(N_c,N_f)$.
Notice that these matrices
are obviously isomorphic to their transpose.
Taking the above step conversely by using
the $(N_f-N_c) \times N_c$ matrices
$h_{\dot{A}}^{T}$,
we will obtain a point of
${\cal M}^{alg}_{\rm b}(N_f-N_c,N_f)$.
Clearly this establishes the isomorphism :
\be
{\cal M}^{alg}_{\rm b}(N_c,N_f)
\simeq {\cal M}^{alg}_{\rm b}(N_f-N_c,N_f),
\ee
which means that the baryonic branches
of $SU(N_c)$ and $SU(N_f-N_c)$ gauge
theories with $N_f$ flavors are exactly same.

%
%
\section{$M$ Theory Description of $N=2$ SQCD}

\subsection{$N=2$ Supersymmetric Gauge Theory in $M$ Theory}

                      In this section we study
$N=2$ supersymmetric QCD from the $M$ theory viewpoint.
Let us begin by describing a basic
configuration of branes in Type IIA picture.
Suppose $N_c$ D fourbranes with worldvolume
$(x^0,x^1,x^2,x^3,x^6)$ stretch between two NS fivebranes
with worldvolume $(x^0,x^1,x^2,x^3,x^4,x^5)$.
On the common four-dimensions $(x^0,x^1,x^2,x^3)$
of their worldvolumes $N=2$ SUSY is preserved
and there appears a $SU(N_c)$ gauge vector multiplet
\footnote{
One $U(1)$ factor is frozen by fixing a center of positions
of D fourbranes}
as the massless mode of the open string among these
$N_c$ D fourbranes. So this configuration
describes $N=2$ supersymmetric pure $SU(N_c)$ gauge theory.
The gauge coupling constant $g$ is determined by
the distance between the two NS fivebranes in $x^6$-direction,
\be
1/g^2=(x^6_2-x^6_1)/2\lambda,
\ee
where $x^6_{1,2}$ denote the positions
of the first and second NS fivebrane
and $\lambda$ is the string coupling constant.
D fourbranes touching NS fivebranes bend \cite{W1}
NS fivebranes logarithmically.
Intersections of D fourbranes with NS fivebranes
in this configuration
will be handled by $(x^4,x^5)$-plane.
It turns out convenient to introduce a complex coordinate
$v$ in this two-plane,
\be
v =  2(x^4+ix^5)/R,
\ee
where $R$ is a constant which will be clear in the next paragraph.
The large $v$ behavior of the $x^6$-position of the NS fivebrane
stuck by the D fourbranes from its left becomes
\be
x^6=k\sum_{a=1}^{N_c}\ln |v-\phi_a| +\const,
\label{log behavior}
\ee
where $k$ is a positive constant
and $\phi_a$ is the position of the $a$-th D fourbrane
in $v$-plane.
The asymptotic behavior of the NS fivebrane stuck by
the D fourbranes from its right has the same form
as (\ref{log behavior}) but with the opposite sign.

                 One can extend the above configuration to $M$ theory.
As the gauge coupling constant in $N=2$ gauge theory is
complexified to $\tau=\frac{\theta}{2\pi}+\frac{4\pi i}{g^2}$,
the corresponding coordinate in Type IIA theory is also
complexified by adding the eleventh dimension $x^{10}$
of $M$ theory. Let us consider $M$ theory
on ${\bf R}^{10}\times {\bf S}^1$,
where the eleventh dimension $x^{10}$ is compactified to
${\bf S}^1$ with radius $R$. It is also useful to introduce
a complex coordinate $s$ by
\be
s=2(x^6+ix^{10})/R.
\ee
The $M$ theory generalization of (\ref{log behavior}) is \cite{W1}
\be
s=2\sum_{a=1}^{N_c}\ln (v-\phi_a) +\const.
\label{log behavior 2}
\ee
The large $v$ behavior of (\ref{log behavior 2})
gives us the relation
\be
-i\tau\simeq 2N_c\ln v,
\label{asym behavior}
\ee
where $-i\tau \equiv (s_2-s_1)/2$.
$s_{1,2}$ are the positions of the
first and second NS fivebranes in $M$ theory.
Then we identify (\ref{asym behavior}) with
the asymptotic behavior of the gauge coupling constant
of $N=2$ $SU(N_c)$ gauge theory.
($-2N_c$ is the coefficient of the one-loop beta function
of the standard asymptotic free (AF) theory,
$-i\tau\simeq -b_0\ln v$.)

                           While eq.(\ref{log behavior}),
obtained in Type IIA picture, is owing to the effect of
the $N_c$ D fourbranes sticking to the fivebranes,
a rationale of eq.(\ref{log behavior 2}) in $M$ theory
is rather profound. These $N_c$ D fourbranes in Type IIA
theory are replaced \cite{W1} by a single $M$ theory
fivebrane with worldvolume ${\bf R}^4\times \Sigma_0$
where $\Sigma_0$ is a Riemann surface of genus $N_c-1$.
This Riemann surface, embedded in the four-dimensional space
$(x^4,x^5,x^6,x^{10})$, which we call $Q_0$,
is identified \cite{W1} with the Seiberg-Witten
curve \cite{SW1,SW2} of pure $SU(N_c)$ gauge theory \cite{KLY,AF}
\be
t^2-2\prod_{a=1}^{N_c}(v-\phi_a)t+\Lambda^{2N_c}=0.
\ee
$t$ relates to $s$ by $t=\exp(-s/2)$ in virtue of the asymptotic
behavior (\ref{asym behavior}).

               An inclusion of matter multiplets into the
above gauge theory can be achieved by putting
$N_f$ D sixbranes with worldvolume
$(x^0,x^1,x^2,x^3,x^7,x^8,x^9)$ between
two NS fivebranes in the basic configuration.
The new configuration obtained thereby describes
$N=2$ $SU(N_c)$ supersymmetric QCD
with $N_f$ flavors.
This is because the open string sector
between $N_c$ D fourbranes and $N_f$ D sixbranes
of the configuration gives the hypermultiplets
which belong to the fundamental representations
both of the gauge and flavor group.

                 These D sixbranes are magnetically
charged with respect to the $U(1)$ gauge field
associated with the circle ${\bf S}^1$
in Type IIA. They behave as the ``Kaluza-Klein
monopoles'' in the four-dimensional space $Q_0$.
Actually they transmute \cite{Townsend} 
$Q_0={\bf R}^3 \times {\bf S}^1$ into
the multi Taub-NUT space, which we denote by $Q$.
Due to the sixbranes the worldvolume of
the aforementioned $M$ theory fivebrane
also changes \cite{W1} to ${\bf R}^4 \times \Sigma$,
where $\Sigma$ is the Seiberg-Witten
curve of $N=2$ supersymmetric QCD with $N_f$ flavors
\cite{HO,APSh}
\be
y^2-2\prod_{a=1}^{N_c}(v-\phi_a)y
+\Lambda^{2N_c-N_f}\prod_{i=1}^{N_f}(v-e_i)=0,
\label{AF curve}
\ee
Notice that
$e_i$ are the bare masses of the matter hypermultiplets
and are identified with the positions of the sixbranes
in $v$-plane. Now the Seiberg-Witten curve $\Sigma$ is
embedded in the multi Taub-NUT space $Q$.

\subsection{Description of the Embedding}

                  In this subsection we provide some
detailed description of the embedding of the Seiberg-Witten
curve $\Sigma$ into the multi Taub-NUT space $Q$.
The formulae obtained in this subsection turn out
useful in the subsequent discussions.

           We shall first deal with the multi Taub-NUT metric.
Introduce the following coordinates $\rv$ and $\sigma$
in the four-dimensional space $(x^4,x^5,x^6,x^{10})$
\be
(\rv,\sigma)=(2x^4/R,2x^5/R,2x^6/R,2x^{10}/R) .
\ee
In terms of these coordinates
the multi Taub-NUT metric on $Q$ will acquire the standard
form \cite{Hawking} :
\be
ds^2=\frac{V}{4}d{\rv}^2+\frac{1}{4V}(d\sigma+\w\cdot d\rv)^2,
\label{mTN}
\ee
where
\bea
&&V=1+\sum_{i=1}^{N_f}\frac{1}{|\rv-\rv_i|}, \nonumber \\
&&\vec{\nabla}\times\w=\vec{\nabla}V
\eea
and $\rv_i$ are the positions of the monopoles
({\it i.e.} sixbranes).
Since $Q$ is a hyper-K\"ahler manifold it will have
the complex structure which fits the Seiberg-Witten curve
$\Sigma$. With such a complex structure one may expect that
the multi Taub-NUT metric becomes K\"ahler.
In order to describe it
let us separate $\rv$ into two parts $v$ and $b$
\be
v=2(x^4+ix^5)/R,~~~~~b=2x^6/R.~~~~~
\left( \sigma=2x^{10}/R. \right)
\ee
By using these variables  metric (\ref{mTN})
acquires the form
\footnote{
We derive expression (\ref{Kahler metric})
of the multi Taub-NUT metric by applying
the technique investigated by
Hitchin \cite{Hitchin}.}
\be
ds^2=\frac{V}{4}dvd\bar{v}
+\frac{1}{4V}\left(\frac{2dy}{y}-\delta dv\right)
\overline{\left(\frac{2dy}{y}-\delta dv\right)}
\label{Kahler metric}
\ee
with
\bea
V &=&
1+\sum_{i=1}^{N_f}\frac{1}{\Delta_i},\\
y &=&
\lambda e^{-(b+i\sigma)/2}
\prod_{i=1}^{N_f}(-b+b_i+\Delta_i)^{1/2},
\label{y in mTN}\\
\delta
&=& \sum_{i=1}^{N_f}
\frac{1}{\Delta_i}\frac{b-b_i+\Delta_i}{v-e_i},\\
\Delta_i
&=& \sqrt{(b-b_i)^2+|v-e_i|^2},
\eea
where $\lambda$ is a constant and $(e_i,b_i)$ represents
the position of the $i$-th monopole.
While $y$ in eq.(\ref{y in mTN}) is determined rather explicitly 
by $v,b$ and $\sigma$,
one can also regard that $v$ and $y$ give the holomorphic
coordinates of $Q$.
With this complex structure the multi Taub-NUT metric
becomes K\"ahler as is clear from (\ref{Kahler metric}).
The K\"ahler form $\omega$ is given by
\be
\omega=i\frac{V}{4}dv\wedge d\bar{v}
+\frac{i}{4V}\left(\frac{2dy}{y}-\delta dv\right)\wedge
\overline{\left(\frac{2dy}{y}-\delta dv\right)},
\label{Kahler form}
\ee
and the holomorphic two-form $\Omega$ is
\be
\Omega=\frac{1}{2}dv\wedge\frac{dy}{y},
\ee
which satisfies the relation,
$\frac{1}{2}\omega\wedge\omega=\Omega\wedge\bar{\Omega}$.

\begin{figure}[t]
\epsfysize=5cm \centerline{\epsfbox{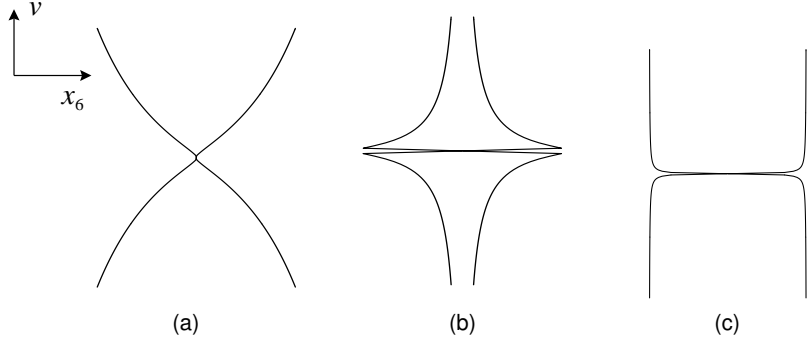}}
\caption{\small
Sections of the Seiberg-Witten curves $\Sigma$ in $M$ theory.
(a) is for asymptotically free theory ($N_f < 2N_c$),
(b) is for IR free theory ($N_f > 2N_c$) and
(c) is for finite theory ($N_f = 2N_c$).
The distance of two NS fivebranes at large $v$ is
infinite, zero and finite, respectively.}
\label{curve in M}
\end{figure}

                       The Seiberg-Witten curve $\Sigma$
given in eq.(\ref{AF curve}) is embedded into $Q$
by using the holomorphic coordinates $(v,y)$
described above. Since $y$ is related with
$b$ and $\sigma$ ($or$ $x^6$ and $x^{10}$) by eq.(\ref{y in mTN})
it makes possible to describe the embedding of the curve
$\Sigma$, say, in $(v,x^6)$-plane.
Some cases are sketched in Fig.\ref{curve in M}.

         In stead of $y$ in (\ref{y in mTN}) one can also take another
choice of the holomorphic coordinates of $Q$ which corresponds to
another branch in curve (\ref{AF curve}).
It is given by $(v,z)$, where $z$ is introduced by
\be
z =
\lambda' e^{(b+i\sigma)/2}
\prod_{i=1}^{N_f}
\frac{v-e_i}{|v-e_i|}
(b-b_i+\Delta_i)^{1/2},
\label{z in mTN}
\ee
which satisfies
\be
yz=\Lambda^{2N_c-N_f}\prod_{i=1}^{N_f}(v-e_i).
\label{yz=v}
\ee

                    Eq. (\ref{yz=v}) shows
that $Q$ describes a resolution of
$A_{N_f-1}$ simple singularity, $yz=v^{N_f}$.
It is resolved by a chain of
$(N_f-1)$ holomorphic 2-spheres.
Each 2-sphere intersects the next at the position
of one of these $N_f$ D sixbranes.
Let us denote a 2-sphere between two sixbranes
$(e_i,b_i)$ and $(e_{i+1},b_{i+1})$ by
$C_i$. The integrals of the two-forms
$\omega$ and $\Omega$ on
$C_i$ become
\bea
\int_{C_i}\Omega &=& 2\pi i(e_{i+1}-e_{i}),\\
\int_{C_i}\omega &=& -2\pi i(b_{i+1}-b_{i}),
\eea
which give the difference between the positions
of these two sixbranes.

          Now we have two descriptions of the embedding
of the Seiberg-Witten curve $\Sigma$ into the multi
Taub-NUT space $Q$.
One is an algebraic geometrical description.
By considering $Q$ as algebraic surface (\ref{yz=v}),
that is,
forgetting the Riemannian structure of $Q$,
the embedding can be described by the set of equations
with respect to $y,z$ and $v$
\bea
\left\{
  \begin{array}{l}
    y+z=2\prod_{a=1}^{N_c}(v-\phi_a), \\
    \\
    yz=\Lambda^{2N_c-N_f}\prod_{i=1}^{N_f}(v-e_i).
  \end{array}
\right.
\eea
The other is a differential geometric
description. By considering $y$ and $z$ as functions
of $b,\sigma$ and $v$ given in (\ref{y in mTN})
and (\ref{z in mTN}), the embedding can be described
by the equation with respect to
$b,\sigma$ and $v$
\bea
y(v,b,\sigma)+z(v,b,\sigma)
=2\prod_{a=1}^{N_c}(v-\phi_a),
\label{y+z=v}
\eea
where
\bea
y(v,b,\sigma) &=&
\lambda e^{-(b+i\sigma)/2}
\prod_{i=1}^{N_f}(-b+b_i+\Delta_i)^{1/2},
\label{y in mTN 2} \\
z(v,b,\sigma) &=&
\lambda' e^{(b+i\sigma)/2}
\prod_{i=1}^{N_f}
\frac{v-e_i}{|v-e_i|}
(b-b_i+\Delta_i)^{1/2}.
\label{z in mTN 2}
\eea

\subsection{The Non-Baryonic Branch}

      In this subsection we study how the non-baryonic branches
are realized in $M$ theory.
To enter the Higgs branch from the Coulomb branch of
$N=2$ $SU(N_c)$ supersymmetric QCD with $N_f$ flavors,
all the bare masses $e_i$ of the matter multiplets
must be same. We will take vanishing bare masses
${}^{\forall} e_i=0$ for simplicity.

       Consider the $r$-th non-baryonic branch root.
The non-baryonic branch and the Coulomb branch come in contact
with each other at this root.
In Type IIA picture $r$ pieces of $N_c$ D fourbranes
are located at the origin of $v$-plane while
the positions of $N_f$ D sixbranes are
$(0,b_i)$ in $(v,b)$-plane.
($b=2x^6/R.$)
Their complicated intersections make
this configuration too singular to allow further analysis
in Type IIA theory.
But, with the eye of $M$ theory,
it actually becomes tractable \cite{W1}.
Let us pay attention to the $M$ theory fivebrane
with worldvolume ${\bf R}^4 \times \Sigma$
where $\Sigma$ is the Seiberg-Witten curve.
Near the non-baryonic branch root the Seiberg-Witten
curve is getting pinched and therefore
the worldvolume of the fivebrane is almost singular.
At the non-baryonic branch root,
as soon as the worldvolume of the fivebrane
becomes singular,
this fivebrane divides into parts and
gives rise to $M$ theory fivebranes wrapping
the two-cycles of $Q$
which resolve $A_{N_f-1}$ simple singularity.

          Let us first examine  how many $M$ theory fivebranes
are wrapping a given two-cycle of $Q$. For simplicity we
assume $b_1 < b_2 < \cdots < b_{N_f}$. Two-cycle between
the two sixbranes $(0,b_i)$ and $(0,b_{i+1})$ is denoted by
$C_i$. ($1 \leq i \leq N_f-1$.)
This cycle is realized by a segment between
$(0,b_i)$ and $(0,b_{i+1})$ in $(v,b)$-plane.
We study the embedding of the Seiberg-Witten curve in
the region of $Q$ which satisfies $|v| \ll  |b-b_i|$ 
for  ${}^{\forall} i$.
In this region $y(v,b,\sigma)$ (\ref{y in mTN 2})
and $z(v,b,\sigma)$ (\ref{z in mTN 2}) behave as
\bea
y(v,b,\sigma) & \sim &
\prod_{i=1}^{N_f}
\left(-b+b_i+|b-b_i|+\frac{1}{2}|v|^2/|b-b_i|\right)^{1/2} ,
\nonumber \\
z(v,b,\sigma) & \sim &
\prod_{i=1}^{N_f}
\left(b-b_i+|b-b_i|+\frac{1}{2}|v|^2/|b-b_i|\right)^{1/2} .
\label{y z near bi}
\eea
Notice the following equalities
\bea
-b+b_i+|b-b_i| &=&
\left\{
\ba{lcl}
2(-b+b_i)\neq 0 & \mbox{if} & b<b_i \\
0         & \mbox{if} & b>b_i
\ea
\right.,
\nonumber \\
b-b_i+|b-b_i| &=&
\left\{
\ba{lcl}
0 & ~~\mbox{if} & b<b_i \\
2(b-b_i) \neq 0         & ~~\mbox{if} & b>b_i
\ea
\right. .
\nonumber
\eea
Therefore,
in the $i$-th interval between $b_i$ and $b_{i+1}$,
that is, $b_i < b < b_{i+1}$,
eqs. (\ref{y z near bi}) become as follows
\bea
y(v,b,\sigma) &\sim&
v^i \times (\mbox{\it non-zero factor}),
\nonumber \\
z(v,b,\sigma) &\sim&
v^{N_f-i} \times (\mbox{\it non-zero factor}).
\label{y z near bi 2}
\eea
Near $v=0$,
embedding (\ref{y+z=v}) of the Seiberg-Witten curve
can be described approximately  by
\be
y(v,b,\sigma) + z(v,b,\sigma) \sim v^r .
\label{embed at bi}
\ee
Inserting eqs. (\ref{y z near bi 2})
into eq. (\ref{embed at bi})
we can see that the lowest order in $v$ among $y,z$
and $v^r$ is factored out from eq. (\ref{embed at bi})
and counts the multiplicity of zeros.
Therefore this power of $v$ factored out provides
the number of $M$ theory fivebranes wrapping the two-cycle
$C_i$.

\begin{figure}[t]
\epsfysize=5.5cm \centerline{\epsfbox{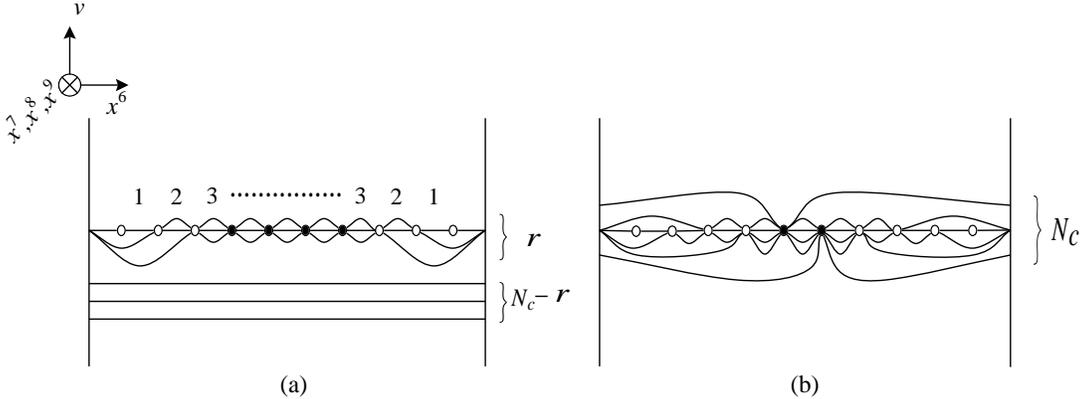}}
\caption{\small (a) $r$-th non-baryonic branch. (b) baryonic branch.}
\label{NBandB}
\end{figure}

                If we denote this  multiplicity
in the $i$-th gap by $n_i$,
there are the following three cases :
\be
n_i=\left\{
\ba{lcl}
i     & \mbox{for} & 1 \leq i < r \\
r     & \mbox{for} & r \leq i \leq N_f-r \\
N_f-i & \mbox{for} & N_f-r < i \leq N_f-1
\ea
\right. .
\label{s-rule}
\ee
This is nothing but the so-called $s$-rule \cite{HW}!
\footnote{The $s$-rule is also proved in
\cite{HOO} from the point of view of
an algebraic singularity resolution.}
With this $s$-rule we can find the correct
brane configuration of the $r$-th non-baryonic
branch root. It is given in Fig.\ref{NBandB}(a).
Each $M$ theory fivebrane wrapping the two-cycle
$C_i$ becomes, in Type IIA picture,
a D fourbrane which worldvolume coincides with
the segment between $(0,b_i)$ and $(0,b_{i+1})$
in $(v,b)$-plane.

           The $s$-rule (\ref{s-rule})
shows that the total number of $M$ theory fivebranes
wrapping the vanishing cycles of $Q$ is
$r(N_f-r)$.
These fivebranes (or D fourbranes in Type IIA picture)
provide \cite{W1} massless hypermultiplets.
Actually, for each 2-sphere $C$ we can associate
four real parameters:
Its position $\vec{w}=(w^1,w^2,w^3)$ in
$(x^7,x^8,x^9)$-directions and 
$w^4=i\int_{C}\omega$, where $\omega$ is
the K\"ahler form (\ref{Kahler form}).
These four parameters provide the vev of
one hypermultiplet $q_{\dot{A}}$ by
$q_{\dot{1}} = w^1 + iw^2$ and
$q_{\dot{2}} = w^3 + iw^4$.

\begin{figure}[t]
\epsfysize=6cm \centerline{\epsfbox{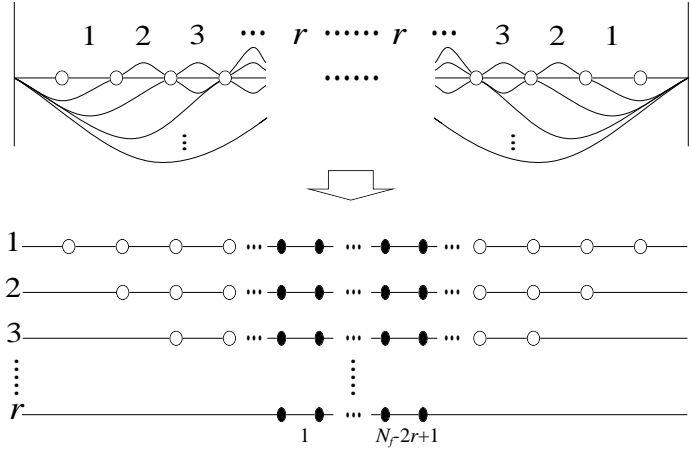}}
\caption{\small Resolution of degenerated D fourbranes.}
\label{GlobalSym}
\end{figure}

          The moduli space
${\cal M}_{\bf nb}^{(r)}$
of the $r$-th non-baryonic branch should be
described by these $r(N_f-r)$ hypermultiplets.
To confirm it let us consider the residual global symmetry
of the brane configuration, that is, the symmetry which does not
move the brane configuration.
We start with a phenomenological argument.
Consider the resolution of the degenerated
$M$ theory fivebranes or D fourbranes as sketched in
Fig.\ref{GlobalSym}.
There appear $r$ chains of the $M$ theory fivebranes.
The $i$-th chain consists of $(N_f-2i+1)$ 2-spheres, 
which give $(N_f-2i+1)$ hypermultiplets
${\bf q}_{\dot{A}}^i=
(q_{\dot{A}}^{i1}, \cdots, q_{\dot{A}}^{i N_f-2i+1})$.
$1 \leq i \leq r$.
We shall pay attention to the $r$-th chain first.
By using $SU(2)_R$ and a subgroup
$U(N_f-2r+1)$ of the flavor group $U(N_f)$,
one may rotate these $(N_f-2r+1)$ hypermultiplets
${\bf q}_{\dot{A}}^r$ into a normal form,
${\bf q}_{\dot{1}}^r=0$ and
${\bf q}_{\dot{2}}^r=(d_r,0,\cdots,0)$
where $d_r \in {\bf R}_{>0}$.
Then we can see that the residual symmetry of the $r$-th chain
is $U(N_f-2r) \times U(1)$.
The appearance of the extra $U(1)$ symmetry will be shown in the
next paragraph.
Secondly, we consider the $(r-1)$-th chain.
It provides $(N_f-2r+3)$ hypermultiplets
${\bf q}_{\dot{A}}^{r-1}=
(q_{\dot{A}}^{r-1 1}, \cdots, q_{\dot{A}}^{r-1 N_f-2r+3})$.
One may also rotate these hypermultiplets
${\bf q}_{\dot{A}}^{r-1}$, by using $SU(2)_{R}$ and
a subgroup $U(N_f-2r+3)$ of $U(N_f)$, into a normal form,
${\bf q}_{\dot{1}}^{r-1}=0$ and
${\bf q}_{\dot{2}}^{r-1}=(d_{r-1},0,\cdots,0)$.
One may expect that the residual symmetry
of the $(r-1)$-th chain is $U(N_f-2r+2) \times U(1)$.
But, looking at Fig.\ref{GlobalSym},
it is very plausible that
the non-Abelian part $U(N_f-2r+2)$ includes
$U(N_f-2r)$ of the stabilizer of the $r$-th chain.
Therefore we must take
$U(N_f-2r) \times U(1)$ as the residual
symmetry of the $(r-1)$-th chain.
Repeating the same argument to the 1st chain we obtain
$U(N_f-2r) \times U(1)^r$ as the residual symmetry of the
brane configuration. This coincides with
the stabilizer of the $r$-th non-baryonic branch
described in Section 2.

                  The above phenomenological argument will be
justified by using an algebraic description of
${\cal M}_{\bf nb}^{(r)}$. Since we introduce
${\cal M}_{\bf nb}^{(r)}$ as the  stratum of
${\cal M}_{\bf nb}$ specified by the symmetry breaking
pattern, $SU(N_c) \rightarrow SU(N_c-r)$,
or equivalently the condition, $\rk Q_{\dot{A}}=r$
($c.f.$ (\ref{strata of nb})),
we may handle ${\cal M}_{\bf nb}^{(r)}$ by the following
regularized complex symplectic quotient for
$r \times N_f$ complex matrices $H_{\dot{A}}^{(r)}$
\be
{\cal M}_{\bf nb}^{(r) alg}
\equiv
\frac{\mu_{\bf C}^{(r) -1}(0)^s}{GL(r,{\bf C})},
\label{alg nb}
\ee
where
\be
\mu_{\bf C}^{(r) -1}(0)^s
=\left\{H^{(r)}_{\dot{A}}\left|
\ba{l}
\cdot ~
\mu_{\bf C}^{(r)}
\equiv H^{(r)}_{\dot{1}}H^{(r)\dagger}_{\dot{2}}
=0  \\
\cdot ~
\rk H^{(r)}_{\dot{1}}=r
\ea
\right.\right\},
\ee
and $GL(r,{\bf C})$ acts on $H_{\dot{A}}^{(r)}$ by
$(H^{(r)}_{\dot{1}},H_{\dot{2}}^{(r) \dagger})$
$\mapsto$
$(gH^{(r)}_{\dot{1}},H_{\dot{2}}^{(r) \dagger}g^{-1})$
where $g \in GL(r,{\bf C})$.
A simple parametrization of
${\cal M}_{\bf nb}^{(r) alg}$
is given by
\bea
H^{(r)}_{\dot{1}} &=& \left(
\ba{cc}
\phantom{-h^{(r)}_{\dot{2}}h_{\dot{1}}^{(r) \dagger}} & \\
{\bf 1}_{r}  & h^{(r)}_{\dot{1}} \\
               &
\ea
\right),
\label{hr1}
\\
H^{(r)}_{\dot{2}} &=& \left(
\ba{cc}
                                  & \\
-h^{(r)}_{\dot{2}}h_{\dot{1}}^{(r) \dagger} & h^{(r)}_{\dot{2}} \\
                                  &
\ea
\right),
\label{hr2}
\eea
where $h^{(r)}_{\dot{A}}$ are arbitrary
$r \times (N_f-r)$ complex matrices and
will provide the holomorphic coordinates of
${\cal M}^{(r)}_{\rm nb}$.
The flavor group $U(N_f)$,
acting  on $H^{(r)}_{\dot{A}}$
by $H^{(r)}_{\dot{A}} \mapsto H^{(r)}_{\dot{A}}f$
($f \in U(N_f)$),
now transforms $h^{(r)}_{\dot{A}}$ as follows :
\bea
h^{(r)}_{\dot{1}} \mapsto
h^{(r)f}_{\dot{1}} &=&
(A+h^{(r)}_{\dot{1}}C)^{-1}
(B+h^{(r)}_{\dot{1}}D) ,
\nonumber \\
h^{(r)}_{\dot{2}} \mapsto
h^{(r)f}_{\dot{2}} &=&
(A+h^{(r)}_{\dot{1}}C)^{-1}
h^{(r)}_{\dot{2}}(-h^{(r)\dagger}_{\dot{1}}B+D),
\label{flavor action on hr}
\eea
where
$f =
\left(
\begin{array}{cc}
A & B \\
C & D
\end{array}
\right)
\in U(N_f)$.

        One may construct $r \times (N_f-r)$ complex matrices
$h^{(r)}_{\dot{A}}$ by arranging the hypermultiplets
$q_{\dot{A}}^{ij}$ as follows : 
\be
h^{(r)}_{\dot{A}}=
\left(
\begin{array}{cccccc}
q^{1r}_{\dot{A}} & \cdots & \cdots &
                   \cdots & \cdots & q^{1 N_f-1}_{\dot{A}} \\
\vdots  & q^{2 r-1}_{\dot{A}} & \cdots &
                   \cdots & \cdots & q^{2N_f-3}_{\dot{A}} \\
\vdots  & \vdots  & \ddots  &       &    & \vdots  \\
q^{11}_{\dot{A}} & q^{21}_{\dot{A}} & \cdots &
              q^{r1}_{\dot{A}} & \cdots & q_{\dot{A}}^{r N_f-2r+1}
\end{array}
\right).
\ee
In particular the normal form given in the above
phenomenological argument corresponds to
\bea
h^{(r)}_{\dot{1}} &=& 0 ,
\nonumber \\
h^{(r)}_{\dot{2}} &=&
\left(
\begin{array}{ccccccc}
d_1 &     &        &      &  & &  \\
    & d_2 &        &      &  & &  \\
    &     & \ddots &      &  & &   \\
    &     &        & d_r  &  & &
\end{array}
\right) .
\eea
By considering $U(N_f)$-action (\ref{flavor action on hr})
on this form we can also find that the stabilizer is
nothing but $U(N_f-2r) \times U(1)^r$.

\subsection{The Baryonic Branch}

             The $M$ theory realization of the moduli
space of the baryonic branch is similar to the description
of the non-baryonic branch with $r=N_f-N_c$
except that
the remaining $2N_c-N_f$ D fourbranes
(or a part of the original $M$ theory fivebrane)
must be attached to the D sixbranes.
This is because
the colour symmetry is completely broken
at the baryonic branch. The corresponding
brane configuration is given in Fig.\ref{NBandB}(b).
In this figure $2N_c-N_f$ D sixbranes to which these
extra D fourbranes are attached are depicted as
black circles.
$N_c(N_f-N_c)$ hypermultiplets given by
the $M$ theory fivebranes wrapping the vanishing
cycles of $Q$ will parametrize
the hyper-K\"ahler quotient
$\widehat{\cal M}_{\bf b}(\zeta_{\dot{A}\dot{B}})$
(\ref{submoduli}) through its algebraic description
(\ref{algebraic M}).
As we noted in Section 2.1, this moduli space
gives a resolution of
$\cup_{r=0}^{N_f-N_c}{\cal M}_{\bf nb}^{(r)}$.
By comparing the brane configurations of the non-baryonic
and baryonic branches one can observe  that
the extra $2N_c-N_f$ D fourbranes resolve
the non-baryonic branch by suspending the D sixbranes
(the black circles in Fig.\ref{NBandB}(b)).

          Two NS fivebranes can now move in
$(x^7,x^8,x^9)$-directions since all the D fourbranes
are separated by the D sixbranes.
Their relative position in
$(x^7,x^8,x^9)$-directions, possibly accompanied by
their relative position in $x^{10}$-direction,
will also provide a hypermultiplet.
So, adding this hypermultiplet,
we obtain the moduli space ${\cal M}_{\bf b}$ of
the baryonic branch.

%
%

\section{Higgs Branch Root and Duality}

     In this section we shall examine,
from the $M$ theory viewpoint,
the root of baryonic branch,
where the baryonic branch and the Coulomb branch come
in contact with each other.
It is discussed in \cite{APS}
from the field theory viewpoint that
the baryonic branch root is a point where the gauge symmetry
is quantum mechanically enhanced to
$SU(N_f-N_c)\times U(1)^{2N_c-N_f}$
and the underlying field theory is invariant under
the ${\bf Z}_{2N_c-N_f}$ discrete symmetry
(anomaly-free subgroup of $U(1)_R$).
It is also pointed out that
the baryonic branch root admits to have two descriptions
which are dual to each other.
The ``electric'' description can be applied
if one approaches to the root from the baryonic branch side
while the ``magnetic'' description can be used
when one goes to the root from the Coulomb branch side.
Our purpose is to understand the duality
between these two descriptions
in terms of an exchange of NS fivebranes.

      The exchange of the NS fivebranes
in AF theory or IR-free theory seems difficult
to describe.
This is because NS fivebranes do not
have their definite positions
in these theories.
(See Fig.\ref{curve in M}(a) and (b).)
To avoid this difficulty
it might be convenient to embed these theories
into a finite one.
(See Fig.\ref{curve in M}(c).)
The Seiberg-Witten curve of
$N_f=2N_c$ finite theory has the form
\cite{APS,APSh}
\be
y^2-2\prod_{a=1}^{N_c}(v-\phi_a)y
+\lambda_+\lambda_-
\prod_{j=1}^{2N_c}(v+ \lambda_+e_S-e_j)=0,
\label{finite curve}
\ee
where $e_S=\frac{1}{2N_c}\sum e_j$
is the center of the bare masses $e_j$
and $\lambda_{\pm}$ are given by
\be
\lambda_+
= - \frac{2 F(\tau)}{1-2 F(\tau)}~~,~~~
\lambda_-
= - \lambda_+ +2.
\ee
Notice that $F(\tau)$ is the following
automorphic function of the bare coupling constant
$\tau = \frac{\theta}{\pi}+\frac{8 \pi i}{g^2}$
\be
F(\tau)=
16 q \prod_{n=1}^{\infty}
\left(
\frac{1+q^{2n}}{1+q^{2n-1}}
\right)^8
\ee
with $q=e^{\pi i \tau}$.
The modular transforms of $F(\tau)$ are
\cite{WW}
\bea
F(\tau)
\stackrel{T}{\mapsto}
F(\tau +1)
=
\frac{F(\tau)}{F(\tau)-1},
{}~~~~~
F(\tau)
\stackrel{S}{\mapsto}
F(- 1/ \tau )
=
1- F(\tau) ,
\eea
from which we can see that
$F(\tau)$ is invariant
under the action of the congruence subgroup of level 2,
$\Gamma(2)=$
$\left\{
\left(
\left.
\begin{array}{cc}
a & b \\
c & d
\end{array}
\right) \in SL(2, {\bf Z})~
\right|~
b,c ~~\mbox{even}
\right\}$.
The Seiberg-Witten curve (\ref{finite curve}) itself
admits to have a larger symmetry than $\Gamma(2)$.
It is invariant  under the following $T^2$ and
$S$ transforms
\bea
(\lambda_{\pm}(\tau),~e_j,~\phi_a)
&\stackrel{T^2}{\mapsto}&
(\lambda_{\pm}(\tau+2)=\lambda_{\pm}(\tau),~e_j,~\phi_a) ,
\nonumber \\
(\lambda_{\pm}(\tau),~e_j,~\phi_a)
&\stackrel{S}{\mapsto}&
(\lambda_{\pm}(-1/\tau)=\lambda_{\mp}(\tau),~e_j-2e_s,~\phi_a)  .
\eea

       The $M$ theory description of the finite theory tells
that the asymptotic positions of
the two NS fivebranes in $(x^6,x^{10})$-directions
are given by $\ln \lambda_{\pm}$.
In particular their positions in $x^6$-direction
are $\ln | \lambda_{\pm} |$.
Due to the modular property of $F$
the $S$-transform exchanges their asymptotic positions,
$\lambda_{\pm}(-1/\tau)=\lambda_{\mp}(\tau)$.

       Now let us consider the $Z_{2N_c-N_f}$ symmetric
family of the finite theory :
\be
y^2-2v^{N_f-N_c}(v^{2N_c-N_f}-\varphi^{2N_c-N_f})y
+\lambda_+(\tau)\lambda_-(\tau)
v^{N_f}(v^{2N_c-N_f}-M^{2N_c-N_f}(\tau))=0,
\label{electric curve}
\ee
where $\varphi$ is a constant.
And we introduce $M^{2N_c-N_f}(\tau)$
by the following $S$-invariant function
\be
M^{2N_c-N_f}(\tau)=
\frac{\Lambda^{2N_c-N_f}}
{F(\tau)(1-F(\tau))},
\label{scaling}
\ee
where $\Lambda$ is a constant.
The $Z_{2N_c-N_f}$ symmetry is realized in such a way that
curve (\ref{electric curve}) is invariant under the transform,
$(v,y) \mapsto (\omega v,\omega^{N_c} y)$ where
$\omega^{2N_c-N_f}=1$.

            Since $F(\infty)=0$,
eq. (\ref{scaling}) can be regarded as
defining a double scaling at the weak coupling limit
$\tau \rightarrow i \infty$.
In the region $|v| \ll |M|$
the finite curve (\ref{electric curve})
behaves as
\be
y^2-2v^{N_f-N_c}(v^{2N_c-N_f}-\varphi^{2N_c-N_f})y
+4\Lambda^{2N_c-N_f}v^{N_f}=0,
\label{weak electric curve}
\ee
which describes
the AF theory with
colour $SU(N_c)$ and $N_f$ flavor.
$\Lambda$ is now the strong-coupling
scale of this AF theory.
One can approach to the baryonic branch root
by tuning $\varphi$ so that
$\varphi^{2N_c-N_f}=- \Lambda^{2N_c-N_f}$.
At this value of $\varphi$
the discriminant of the above AF theory curve
becomes a perfect square,
which signals an appearance of $2N_c-N_f$
extra massless hypermultiplets.
This is the ``electric'' description of the baryonic
branch root. The baryonic branch of this AF theory 
is realized in the region 
$|v| \ll |M|$ of the brane configuration of 
finite theory (\ref{electric curve}). 
In this scale, being unable to detect the 
extra $2N_c-N_f$ D sixbranes at 
$v=\omega^i M$ 
$(1 \leq i \leq 2N_c-N_f)$, 
the baryonic branch is given by 
Fig.\ref{NBandB}(b) though it is 
embedded in the finite theory.

        What is the brane configuration
of the $Z_{2N_c-N_f}$ symmetric family
(\ref{electric curve})
at the strong-coupling regime?
We begin by examining how two NS fivebranes behave
as the bare coupling constant $g^2$ becomes larger.
Fix $\theta =0$ for simplicity.
Notice that
$F(\tau=\frac{8 \pi i}{g^2})$ is
real and monotonically increasing
with respect to $g^2$.
In particular
$F$ takes $0,1/2$ respectively at $g^2=0,8\pi$.
Therefore the relative distance of the two NS fivebranes
in $x^6$-direction is decreasing as $g^2$ becomes larger
and it satisfies:
\be
\Delta x^6= \ln |\lambda_+ / \lambda_- |
\rightarrow 0~~~~~
(g^2 \rightarrow 8 \pi).
\ee
At $g^2=8 \pi$ ({\it i.e.} $\tau =i$)
these two NS fivebranes are overlapped
in $(v,b)$-plane.
In fact, if one rescales $y$ to $\tilde{y}=(1-2 F)y$,
curve (\ref{electric curve}) acquires the form
\be
\tilde{y}^2
-2(1-2F)v^{N_f-N_c}(v^{2N_c-N_f}-\varphi^{2N_c-N_f})\tilde{y}
-4F(1-F)v^{N_f}(v^{2N_c-N_f}-M^{2N_c-N_f})=0,
\label{near coincidence}
\ee
which shows that $N_c$ D fourbranes becomes irrelevant at
$g^2=8 \pi$.

                         One may expect that
the above overlapping of two NS fivebranes
is evaded by taking some other route
in the upper half plane
where the bare coupling constant $\tau$ lives.
But it is in vain.
Consider a semi-circle
$L=
\{
\tau =ie^{i \alpha}~(-\pi/2 \leq \alpha \leq \pi/2)
\}$
in the upper half plane.
Since $F(\tau)$ satisfies the relations
$\bar{F}(\tau)=F(- \bar{\tau})$ and
$F(-1/\tau)=1-F(\tau)$, it follows that
$\mbox{Re}~F=1/2$ for  ${}^{\forall} \tau \in L$.
This means
$\ln |\lambda_+ / \lambda_- |=0$ on $L$.
Therefore,
for any $\tau \in L$,
the two NS fivebranes are overlapping in
$(v,b)$-plane,
at least asymptotically.

\begin{figure}[t]
\epsfysize=3cm \centerline{\epsfbox{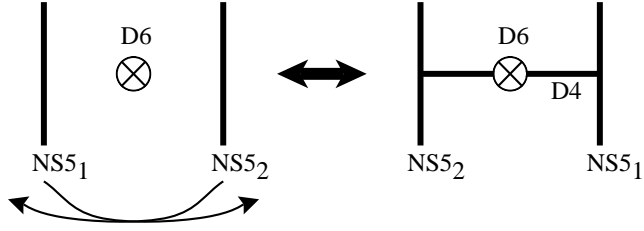}}
\caption{\small
Creation and annihilation of D fourbrane by D sixbrane
in Type IIA picture.}
\label{BraneCreation}
\end{figure}

       The exchange of the two NS fivebranes will take place
on this semi-circle $L$. It is argued in \cite{HW} that,
when two NS fivebranes are exchanged crossing a D sixbrane,
a D fourbrane is created and then suspended between these fivebranes
with touching the sixbrane. Conversely, a D fourbrane suspended
between these fivebranes with touching the sixbrane is annihilated
by this exchange. See Fig.\ref{BraneCreation}.

\begin{figure}[t]
\epsfysize=5cm \centerline{\epsfbox{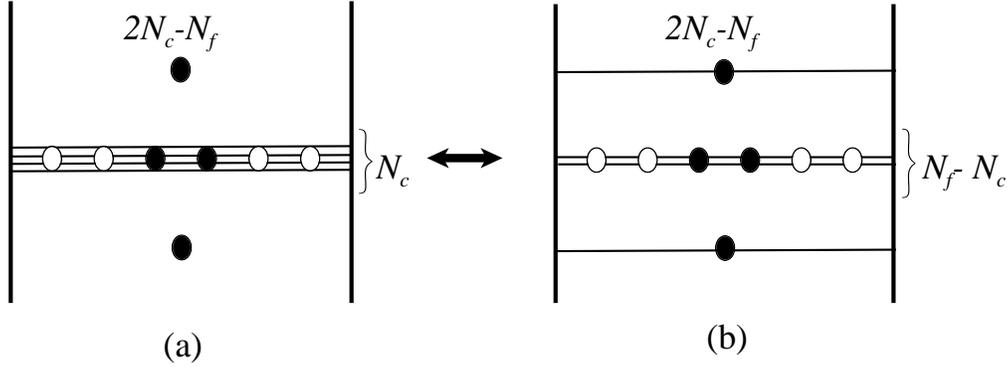}}
\caption{\small
Brane configurations related by
the exchange of the NS fivebranes.
(a) is ``electric'' and (b) is ``magnetic''.
The case of $(N_c,N_f)=(4,6)$ is sketched as
an example.}
\label{duality}
\end{figure}

       With this process of the exchange 
the extra $2N_c-N_f$ D sixbranes 
located at $v=\omega^i M$ 
$(1 \leq i \leq 2N_c-N_f)$ 
create $2N_c-N_f$ D fourbranes there, 
while the $2N_c-N_f$ D fourbranes 
touching the D sixbranes at the origin 
are annihilated. 
Due to the $s$-rule the brane configuration becomes 
a ``magnetic'' configuration depicted in Fig.\ref{duality}.
In this ``magnetic'' configuration (Fig.\ref{duality}(b))
we can find $SU(N_f-N_c)\times U(1)^{2N_c-N_f}$ SQCD
with $N_f$ flavor and
$2N_c-N_f$ massless singlet hypermultiplets
charged by the $U(1)$ factors,
which gives the ``magnetic'' description \cite{APS}
of the baryonic branch root.
Inside the semi-circle $L$
the electric curve (\ref{electric curve}) will
jump to the magnetic one:
\be
y^2-2v^{N_f-N_c}(v^{2N_c-N_f}-M(\tau)^{2N_c-N_f})y
+\lambda_+(\tau)\lambda_-(\tau)
v^{N_f}(v^{2N_c-N_f}-M(\tau)^{2N_c-N_f})=0.
\label{magnetic curve}
\ee
This curve is also invariant under the $S$-transform and
one can study it in terms of the dual bare coupling
constant $\tilde{\tau}=-1/\tau$.

          The $2N_c-N_f$ massless singlet 
hypermultiplets obtained from the open string 
among the extra $2N_c-N_f$ D sixbranes and 
the D fourbranes touching them do not have 
their moduli. Namely they are frozen. 
The net moduli space of the Higgs branch 
of this theory is that of 
$SU(N_f-N_c)$ SQCD with 
$N_f$ flavor. 
As we have seen in Section 2, 
the baryonic branch of this IR-free theory 
is isomorphic to that of 
$SU(N_c)$ SQCD with 
$N_f$ flavor.

%
%

\section*{Acknowledgments}

We would like to thank H. Itoyama and S. Hirano for useful
discussions.
T.N. is supported in part by
Grant-in-Aid for Scientific Research 08304001.
K.O. and Y.Y. are supported in part by the JSPS
Research Fellowships.

%
%

\end{document}